\documentclass{article}

\usepackage{arxiv}

\usepackage[utf8]{inputenc} 
\usepackage[T1]{fontenc}    
\usepackage{hyperref}       
\usepackage{url}            
\usepackage{booktabs}       
\usepackage{amsfonts}       
\usepackage{nicefrac}       
\usepackage{microtype}      
\usepackage{lipsum}		
\usepackage{graphicx}
\usepackage{doi}
\usepackage[version=4]{mhchem}
\usepackage[labelfont=bf]{caption}

\usepackage[
    backend=biber,
    style=chem-angew,
    autocite=superscript
]{biblatex}

\usepackage[usenames,dvipsnames]{xcolor}\usepackage[usenames,dvipsnames]{xcolor}
\newcommand\myshade{85}
\colorlet{mylinkcolor}{violet}
\colorlet{mycitecolor}{NavyBlue}
\colorlet{myurlcolor}{Aquamarine}

\hypersetup{
  linkcolor  = mylinkcolor!\myshade!black,
  citecolor  = mycitecolor!\myshade!black,
  urlcolor   = myurlcolor!\myshade!black,
  colorlinks = true,
}

\renewcommand{\undertitle}{Preprint -- accepted by Chem. Comm. (RSC) on Nov 4, 2022\\ (\textit{Chem. Commun.} \textbf{2022}, \textit{58}, 13369-13372. DOI: \href{https://dx.doi.org/10.1039/D2CC04928G}{10.1039/D2CC04928G})}
\renewcommand{\headeright}{Preprint}

\title{Solving a problem with a single parameter: A smooth \textit{bcc} to \textit{fcc} phase transition for metallic lithium}

\author{ \href{https://orcid.org/0000-0002-2995-9755}{\includegraphics[scale=0.06]{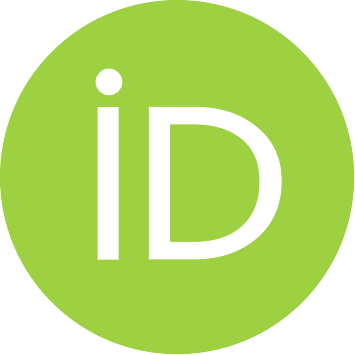}\hspace{1mm}Paul Jerabek$^*$} \\
	Institute of Hydrogen Technology\\
	Helmholtz-Zentrum Hereon\\
	Max-Planck-Str. 1 \\
    21502 Geesthacht (Germany)\\
	\texttt{paul.jerabek@hereon.de} \\
 	\And
	\href{}{Anthony Burrows} \\
	The New Zealand Institute for Advanced Study\\
    Center for Theoretical Chemistry and Physics\\
	Massey University\\
	0632 Auckland (New Zealand) \\
	\And
	\href{https://orcid.org/0000-0003-4845-686X}{\includegraphics[scale=0.06]{orcid.pdf}\hspace{1mm}Peter Schwerdtfeger$^*$} \\
	The New Zealand Institute for Advanced Study\\
    Center for Theoretical Chemistry and Physics\\
	Massey University\\
	0632 Auckland (New Zealand) \\
	\texttt{peter.schwerdtfeger@gmail.com} \\
}

\renewcommand{\shorttitle}{A smooth \textit{bcc} to \textit{fcc} phase transition for metallic Lithium}

\hypersetup{
pdftitle={Solving a problem with a single parameter: A smooth bcc to fcc phase transition for metallic lithium},
pdfsubject={},
pdfauthor={Paul Jerabek, Anthony Burrows, Peter Schwerdtfeger},
pdfkeywords={Martensitic Transformation, Lithium Metal, Density
Functional Theory, Phase Transition Path, Stability of the Body-Centred Cubic Phase},
}

\begin{document}
\maketitle

\begin{abstract}
Density functional calculations for metallic lithium along a cuboidal \textit{bcc}-to-\textit{fcc} transformation path demonstrate that the \textit{bcc} phase is quasi-degenerate with the \textit{fcc} phase with a very small activation barrier of 0.1~kJ/mol, but becomes the dominant phase at higher temperatures in accordance with Landau theory. This resolves the long-standing controversy about the two phases for lithium.
\end{abstract}

\keywords{Martensitic Transformation \and Lithium Metal \and Density
Functional Theory \and Phase Transition Path \and Stability of the Body-Centred Cubic Phase}

Lithium is the lightest metal in the Periodic Table, and a comparatively rare but thought-after element for many applications in industry such as for lithium battery technologies.\autocite{wang2015a} It crystallizes in the body-centred cubic (\textit{bcc}) phase (at normal conditions) accompanied by energetically close-by face-centred cubic (\textit{fcc}) and martensitic Sm-type \textit{9R} phases (stacking sequence ABCBCACAB),\autocite{overhauser1984a,smith1987a,ashcroft1989a,schwarz1990a,ackland2017a,smith1990a} the latter can be interpreted as a Barlow packing\autocite{barlow1883a} in-between \textit{fcc} and \textit{hcp} (hexagonal close-packed) stacking arrangements.\autocite{berliner1986a} Lithium shows very unusual behavior in its chemical and physical properties, e.g. at pressures above 20~GPa one observes a superconducting phase,\autocite{ashcroft2002a,schaeffer2015a,struzhkin2002a,shimizu2002a} above 40~GPa a liquid phase appears at room temperature,\autocite{guillaume2011a} and at higher pressures a metal-to-semi-conductor/insulator transition occurs\autocite{matsuoka2009a} and electron pairing is predicted.\autocite{neaton1999a}

\begin{figure}[b!]
\centering
  \includegraphics[width=.75\columnwidth]{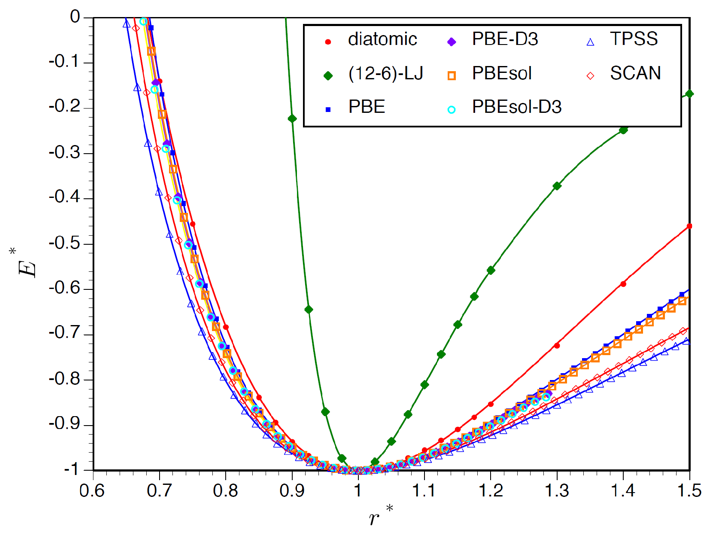}
  \caption{Potential energy curves $E^*(r^*)$ for \ce{Li2} at different DFT levels in comparison to a (12,6)-LJ potential. Dimensionless units are used, i.e. $E^*(r^*) = \varepsilon^{-1} E(r/r_{\mathrm{min}})$, with the experimental dissociation energy $\varepsilon$ = 1.056~eV and equilibrium distance $r_{\mathrm{min}}$ = 2.673~\AA~ for the LJ potential.\autocite{barakat1986a} The (12-6) LJ scaled potential for the diatomic and for the cohesive energy of the solid have identical functional forms.}
  \label{fig:potential-energy-curve}
\end{figure}
There has been a long-standing debate about the stability and existence of the body-centred cubic (\textit{bcc}) phase observed for a few metals in the Periodic Table. For such elements, the transition from \textit{bcc} to the closest packed structures, termed martensitic transformation, involves rather complex rearrangements of the atoms in the lattice and has therefore remained elusive for experimental observation. 
Similarly, the exact mechanism of the \textit{bcc}$\rightarrow$\textit{fcc}/\textit{hcp}/\textit{9R} martensitic phase transition has also been a matter of debate.\autocite{kraft1993a,rollmann2007a,cayron2015a,pichl1999a,hecht2020a} Barrett in 1956 described for the first time the co-existence of the \textit{bcc} and close packed phases for lithium at 78~K,\autocite{barrett1956a} with the \textit{bcc} phase becoming dominant at higher temperatures. The transition temperature rises with higher pressures for lithium.\autocite{vaks1989a} Recent studies indicate that the martensitic \textit{bcc}$\rightarrow$\textit{fcc} transformation is a diffusionless phase transition,\autocite{ahlers2004a} i.e. a homogenous distortion where atoms move in a concerted fashion with small changes in distances and volumes along the transformation path (although the \textit{bcc}$\rightarrow$\textit{9R} transition might be defect induced\autocite{schwarz1991a}). In such a phase transition one goes through lattice structures which can be interpreted as mean geometric lattices between the starting and ending points.\autocite{barrett1956a} Conway and Sloane introduced such a lattice, the mean centred cuboidal lattice (\textit{mcc}), which is the densest isodual lattice of all cuboidal lattices and seen as a combination between the \textit{bcc} and \textit{fcc} lattice.\autocite{conway1994a}

A remarkable property of lithium is that the gas phase dimer has an extremely broad potential energy curve in both the short- and long-range regions,\autocite{barakat1986a} thus deviating substantially from an ideal Lennard-Jones (LJ) system.\autocite{burrows2021a} This feature propagates into the solid state for the cohesive energy as Fig. \ref{fig:potential-energy-curve} shows.
For the simple (12,6)-LJ potential the \textit{bcc} phase represents a maximum on the potential energy surface, and transforms into the \textit{fcc} phase without a barrier.\autocite{burrows2021a} In contrast, X-ray diffraction studies for lithium show that the most stable phase at low temperatures is \textit{fcc}, with the \textit{bcc} and \textit{9R} phases most likely being metastable.\autocite{overhauser1984a} 
Here we adopt the idea of Conway and Sloane and introduce lattice vectors for the primitive cuboidal cell, dependent on one single parameter $A$ beside the nearest neighbour (NN) distance $r$ between the atoms,\autocite{burrows2021a}
\begin{equation}
 \begin{split}
 \vec{b}^{\,T}_{1} &= r(1,0,0); \quad \vec{b}^{\,T}_{2} = \frac{r}{A+1} \bigg( A,\, \sqrt{2A+1},\, 0  \bigg); \\
 \vec{b}^{\,T}_{3} &= \frac{r}{(A+1)\sqrt{2A+1}} \bigg( \sqrt{2A+1},\, 1,\, \sqrt{4A(A+1)}\, \bigg)
 \end{split}
\end{equation}
\begin{figure}[htb!]
\centering
 \begin{tabular}{cc}
  \includegraphics[width=0.45\columnwidth]{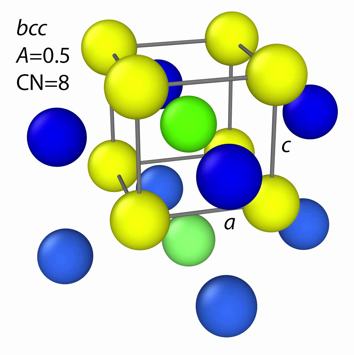} \hfill & \includegraphics[width=0.45\columnwidth]{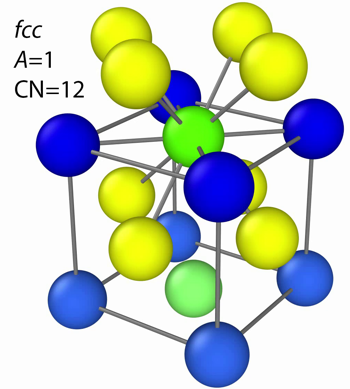} \\
  (a)  & (b) 
 \end{tabular}
  \caption{The \textit{bcc} and \textit{fcc} lattices along the cuboidal transformation path. CN is the coordination number. (a) \textit{bcc} ($A = 1/2$) cell: the 8 yellow atoms define the conventional \textit{bct} cell with lattice parameters ($a,c$). (b) \textit{fcc} ($A = 1$): the 8 blue, 4 lower yellow and 2 green atoms define the conventional \textit{fcc} cell. 
  }
  \label{fig:lattices}
\end{figure}

Through the parameter $A$ we smoothly connect the axially-centred cuboidal \textit{acc} lattice ($A=1/3$) with the \textit{bcc} lattice ($A=1/2$), the \textit{mcc} lattice ($A=1/\sqrt{2}$) and finally the \textit{fcc} lattice ($A=1$). These cuboidal cells along the martensitic transition path are of Bain-type\autocite{bain1924a} and belong all to the common space-group \textit{I4/mmm} of a body-centred tetragonal cell (\textit{bct}), see Fig. \ref{fig:lattices}.

A number of different density functionals were tested for lattice optimizations to obtain cohesive energies and NN distances dependent on the parameter $A$. They all give very similar results when \textit{fcc} is taken as a reference (see ESI). However, cohesive energies can vary widely from $-$148 (PBE0) to $-$177~kJ/mol (PBEsol-D3) (see also Doll et al.\autocite{doll1999a} or Faglioni et al.\autocite{faglioni2016a}). We decided to choose the PBE functional for all further investigations, as it gives excellent results compared to experiment\autocite{berliner1986a,felice1977a,anderson1985a}(Tab. \ref{tab:solid-state}) or quantum Monte-Carlo results.\autocite{yao1996a} According to our calculations, the \textit{fcc} lattice is slightly preferred over the \textit{hcp} lattice by 0.015~kJ/mol at the PBE level of theory. Such a tiny energy difference between \textit{fcc} and \textit{hcp} explains perhaps why Barlow packings\autocite{barlow1883a,hopkins2011a} such as the \textit{9R} phase have been observed for lithium.\autocite{overhauser1984a}

\begin{table}[htb!]
\small
  \caption{\ Solid state properties (at 0~K) at the PBE level of theory with and without zero-point vibrational energies (ZPVE) included for the lattices \textit{fcc} and \textit{bcc} and for the isotopes \ce{^6Li} and \ce{^7Li}.}
  \label{tab:solid-state}
  \begin{tabular}{lcccc}
    \hline
                          &      PBE      &    PBE+ZPVE   &    PBE+ZPVE   &      Exp.\textsuperscript{a}\\
                          &               &    \ce{^6Li}  &    \ce{^7Li}  &      \ce{^7Li}  \\
    \hline
     \textit{fcc}($A=1$)  &               &               &               &                 \\
     $r_{\mathrm{min}}$   &   3.0570      &   3.0749      &   3.0754      &    3.111        \\
     $V$                  &   20.199	  &   20.555	  &   20.568	  &    21.3(2)      \\
     $\rho$               &   0.5755\textsuperscript{b}	  &   0.5655	  &   0.5664	  &    0.5471       \\
     $E_{\mathrm{coh}}$  &   --155.24	  &   --151.33	  &   --151.29	  &    --158        \\
     $B$                  &   13.94  	  &   --     	  &   --    	  &    12.46(17)    \\
                          &          	  &           	  &          	  &    12.95        \\
    \hline
    \textit{bcc}($A=1/2$) &               &               &               &                 \\
     $r_{\mathrm{min}}$   &   2.9734      &   2.9969      &   2.9951      &                 \\
     $V$                  &   20.236	  &   20.720	  &   20.682	  &                 \\
     $\rho$               &   0.5744\textsuperscript{b}	  &   0.5601	  &   0.5620	  &                 \\
     $E_{\mathrm{coh}}$  &   --155.09	  &   --150.99	  &   --151.53	  &                 \\
     $B$                  &   13.82 	  &   --     	  &   --    	  &                 \\
    \hline
           TS\textsuperscript{c}&                &               &               &         \\
            $A_{\mathrm{TS}}$   &   0.646    	  &   0.680  	  &   0.679  	  &         \\
     $\Delta E_{\mathrm{TS}}$   &   0.185 	      &   0.112    	  &   0.116   	  &         \\
    \hline
  \end{tabular}
  
  \footnotesize Optimized equilibrium nearest neighbor distance $r_{\mathrm{min}}$ in \AA, corresponding volumes $V$ in \AA\textsuperscript{3}, density $\rho$ in g/cm\textsuperscript{3}, cohesive energy $E_{\mathrm{coh}}$ in kJ/mol and bulk modulus $B$ in GPa. \textsuperscript{a}Experimental values from refs. \autocite{overhauser1984a,felice1977a,anderson1985a}. \textsuperscript{b}The mass of \ce{^7Li} is taken for the density. \textsuperscript{c}The location of the transition state (TS) is defined through the parameter $A$ with the activation energy, $\Delta E_{\mathrm{TS}} = E(A_\mathrm{TS}) - E(A=1)$, in kJ/mol.
  
\end{table}

\begin{figure}[htb!]
\centering
\includegraphics[width=\columnwidth]{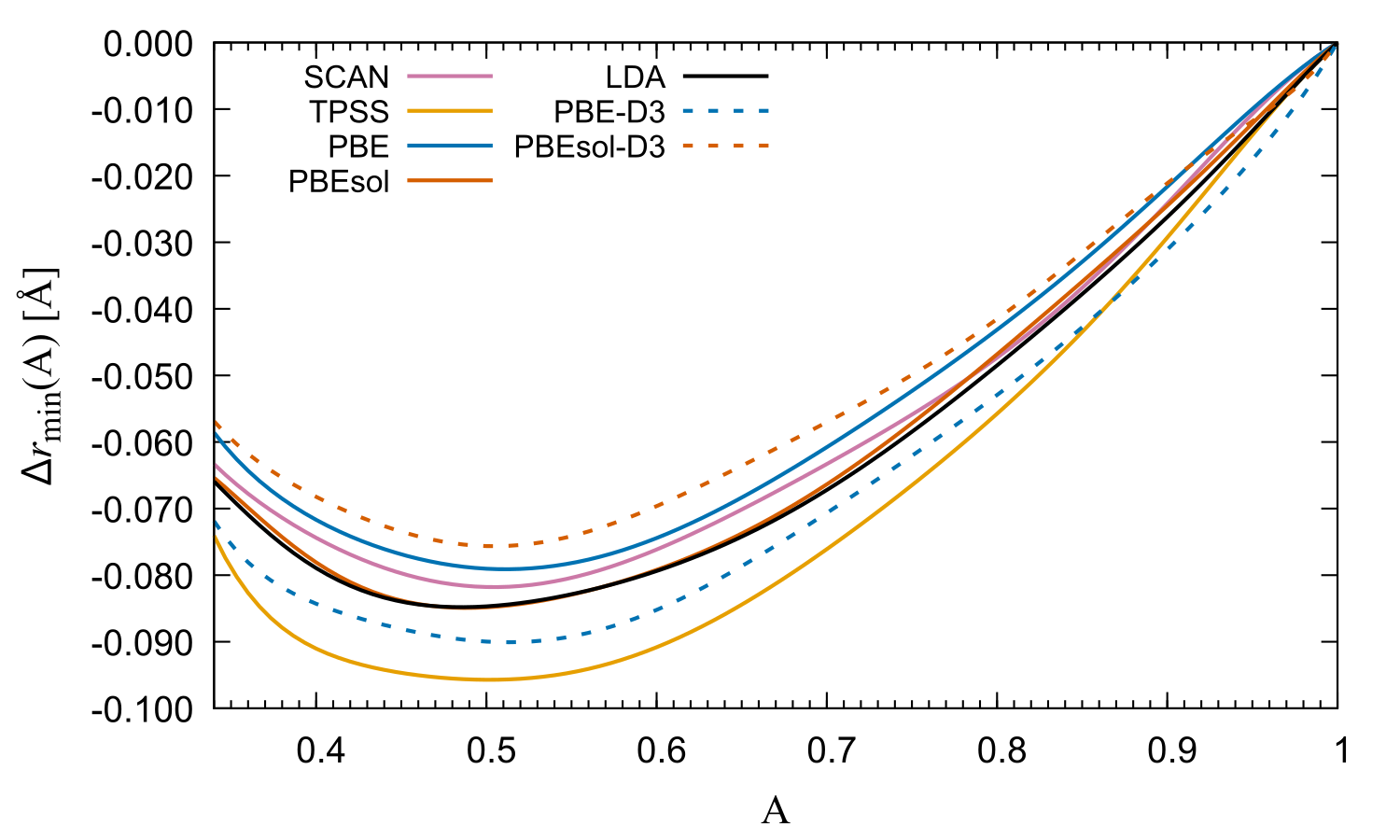}
  \caption{Differences in optimized NN distances $\Delta r_{\mathrm{min}}(A) = r_{\mathrm{min}}(A) - r_{\mathrm{min}}(\mathrm{\textit{fcc}})$  at different levels of DFT.}
  \label{fig:r-vs-a_rel-min}
\end{figure}

Concerning the martensitic transformation path in the chosen parameter range $1/3 \leq A \leq 1$, we relate the primitive lattice to the \textit{bct} conventional cell with lattice parameters ($a,c$) (Fig. \ref{fig:lattices}),
\begin{eqnarray}
 r = \frac{a}{2}\sqrt{2 + \bigg( \frac{c}{a} \bigg)^2}; \,\,\, A=\frac{1}{2} \bigg( \frac{c}{a} \bigg)^2
\end{eqnarray}
As the parameter $A$ decreases, the ratio $c/a$ decreases and for our parameter range we have $\sqrt{2/3} \leq c/a \leq \sqrt{2}$. Because the change in the NN distance $r$ is less pronounced in this transformation (Fig. \ref{fig:r-vs-a_rel-min}), it makes the selection of the parameters ($r,A$) for the primitive cell far more convenient than ($a,c$) for the conventional cell, where typically for the latter two-dimensional plots are used.\autocite{koumatos2016a} Here the cell parameters ($a,c$) change from (in \AA) (3.057, 4.323)$\rightarrow$(3.433, 3.433)$\rightarrow$(3.672, 2.998) for the \textit{fcc}$\rightarrow$\textit{bcc}$\rightarrow$\textit{acc} transformation. This corresponds to a Bain strain.\autocite{koumatos2016a} For better illustration of the collective movement of atoms in the Bain transformation long the cuboidal transformation path we provide a movie in the ESI. The \textit{fcc}$\leftrightarrow$\textit{hcp} transformation has been suggested to be far more complicated involving diffusion, stacking faults and defects.\autocite{li2017a}

In the Bain diffusionless transformation, the \textit{bcc} structure ($A = 1/2$) has the smallest NN distance of all the cuboidal lattices and \textit{fcc} ($A = 1$) the largest, see Fig. \ref{fig:r-vs-a_rel-min}. However, this does not affect the volume much which is dominated by the change in the parameter $A$, i.e.
\begin{eqnarray}
 V(r,A) = \frac{a c^2}{2} = 2r^3 \sqrt{A} (A+1)^{-3/2} 
\end{eqnarray}

Here the \textit{bcc} lattice has the largest volume in the $A$ range considered and therefore the least dense packing density\autocite{dunlap2017a} given by
\begin{eqnarray}
 \rho(A) = \frac{\pi}{12} \sqrt{\frac{(A+1)^3}{A}} .
\end{eqnarray}

\begin{figure}[t]
\centering
\includegraphics[width=.95\columnwidth]{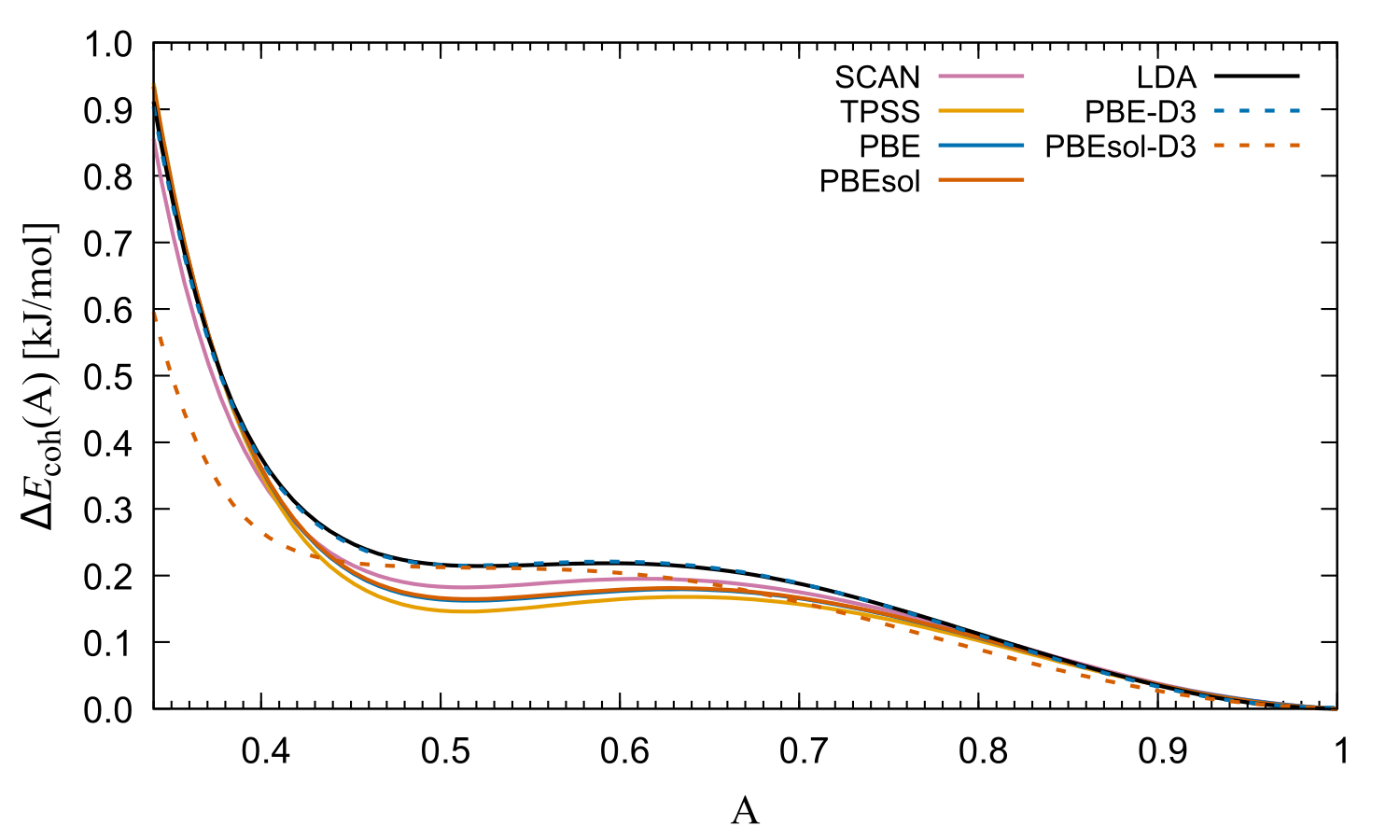}
  \caption{Differences in cohesive energies $\Delta E_{\mathrm{coh}}(A) = E_{\mathrm{coh}}(A) - E_{\mathrm{coh}}(\mathrm{\textit{fcc}})$ at different levels of DFT. $E_{\mathrm{coh}}(\mathrm{\textit{fcc}})$ is strongest for all tested functionals.}
  \label{fig:e-vs-a_rel-min}
\end{figure}

Fig. \ref{fig:e-vs-a_rel-min} depicts the cohesive energies corresponding to each value of $A$ and illustrates that the strongest bonding between Li atoms is obtained for the \textit{fcc} structure ($A = 1$), with the \textit{bcc} structure representing a very shallow metastable minimum in accordance with Xie et al.\autocite{xie2008a} At the PBE level of theory, the transition state (TS) structure located at $A = 0.646$ is close to the \textit{mcc} structure ($A=1/\sqrt{2} = 0.707$). The TS barrier height is merely 0.032~kJ/mol above the \textit{bcc} point (Tab. \ref{tab:solid-state}). This behaviour does not change by using other density functional approximations (Fig. \ref{fig:e-vs-a_rel-min}). The difference between the \textit{bcc} and \textit{fcc} cohesive energies agrees with previous estimates,\autocite{faglioni2016a,liu1999a,hanfland2000a,karasiev2012a,hutcheon2019a} but varies between the different density functionals from 0.14 (TPSS) to 0.40 (HS06) kJ/mol (see ESI). We should be reminded that 0.1~kJ/mol corresponds to a temperature of 12~K and the energy differences between the different structures are very tiny. It is interesting that the \textit{acc} structure ($A=1/3, c<a$) with a coordination number of CN=10 and a higher packing density ($\rho=2\pi/9=0.6981$) than \textit{bcc} (CN=8, $\rho=\pi \sqrt{3}/8=0.6802$) is energetically unstable.
\begin{figure}[htb!]
\centering
\includegraphics[width=\columnwidth]{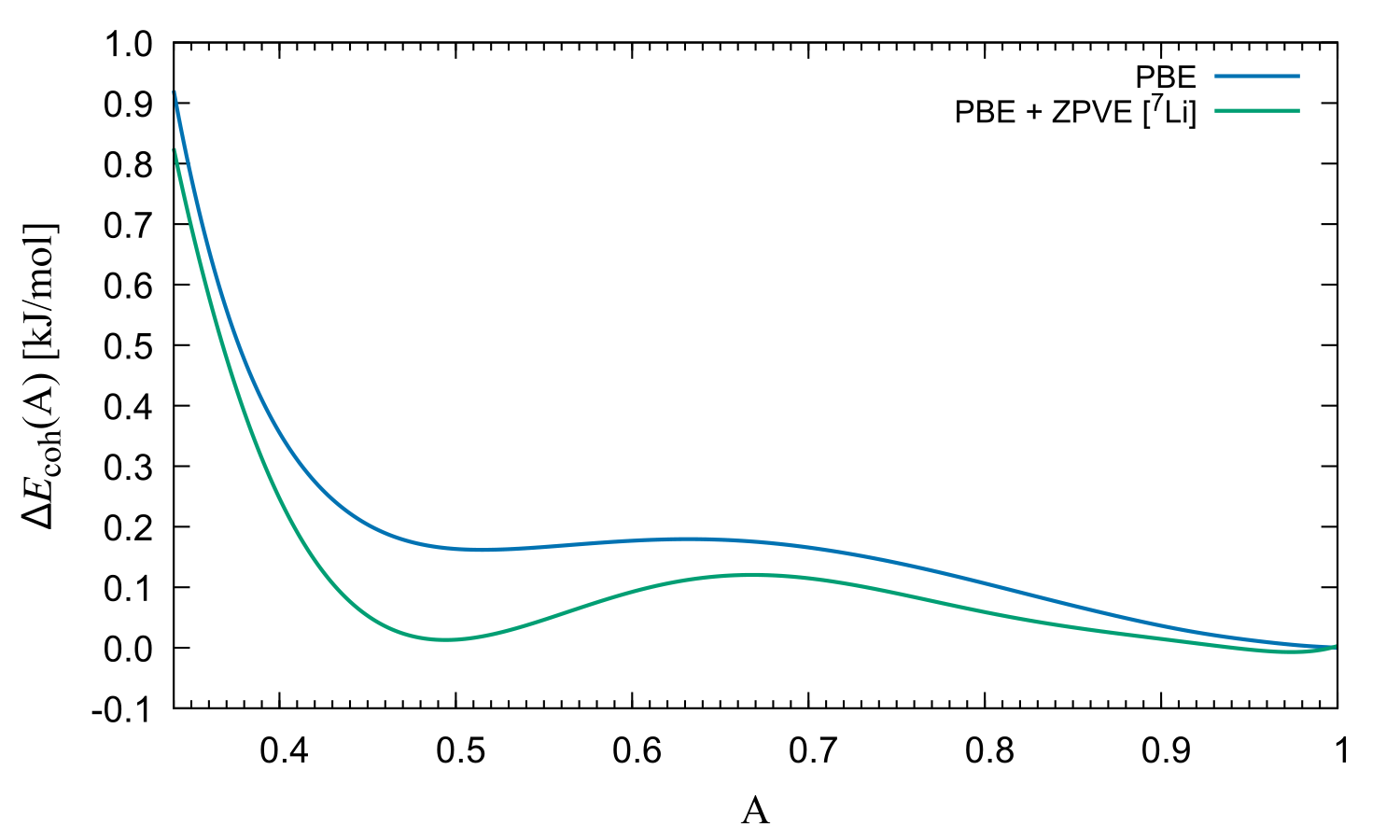}
  \caption{Differences in cohesive energies $\Delta E_{\mathrm{coh}}(A) = E_{\mathrm{coh}}(A) - E_{\mathrm{coh}}(\mathrm{\textit{fcc}})$ for PBE with and without ZPVE contributions obtained via calculation of phonon spectra. At this level of theory, the \textit{bcc} and \textit{fcc} structures ($A = 1/2$ and $1$, respectively) are energetically quasi-degenerate due to the ZPVE contributions. All calculations were carried out at the optimized $A$-dependent structure obtained from the internal (cohesive) DFT energy only.}
  \label{fig:e-vs-a_plot_rel-min_pbe_zpe}
\end{figure}

The picture changes when ZPVEs ($E_\mathrm{ZPVE}$), are included as shown in Fig. \ref{fig:e-vs-a_plot_rel-min_pbe_zpe}. Now the \textit{bcc} structure sits energetically slightly below the \textit{fcc} structure, but the difference is again tiny and will depend on the density functional applied (Tab. \ref{tab:solid-state}). The simple reason is that in the \textit{fcc} lattice an atom has 12 nearest neighbours compared to 8 for \textit{bcc} leading to a larger ZPVE contribution for the \textit{fcc} structure. As a result, the two structures can be considered as quasi-degenerate at this level of theory. The barrier is located more closely ($A=0.68$) to the \textit{mcc} structure with barrier heights of ca. 0.098 and 0.092~kJ/mol above the \textit{bcc} structure for \ce{^6Li} and \ce{^7Li} respectively. This very flat low-energy martensitic path from \textit{bcc} to \textit{fcc} explains the experimental observations and surrounding controversies over the past decades on the correct ground state symmetry for the lithium crystal at low temperatures.\autocite{overhauser1984a,ackland2017a}

Concerning finite temperatures, we discuss the temperature dependent cuboidal transformation through the Helmholtz free energy expression
\begin{eqnarray}
F(T,A) = E_{\mathrm{coh}}(A) + E_{\mathrm{ZPVE}}(A)+E_{\mathrm{therm}}(T,A) - TS(T,A)
\label{eq:free-energy}
\end{eqnarray}
where $E_{\mathrm{therm}}$ arises from the thermal occupation of both phonon and electron states, and $S=S_{\mathrm{vib}}+S_{\mathrm{elec}}$ is the sum of vibrational and electronic contributions to the entropy (for details see ESI). The electronic contributions for the last two terms in Eq. \ref{eq:free-energy} are however small compared to the vibrational contributions.
\begin{figure}[htb!]
\centering
\includegraphics[width=\columnwidth]{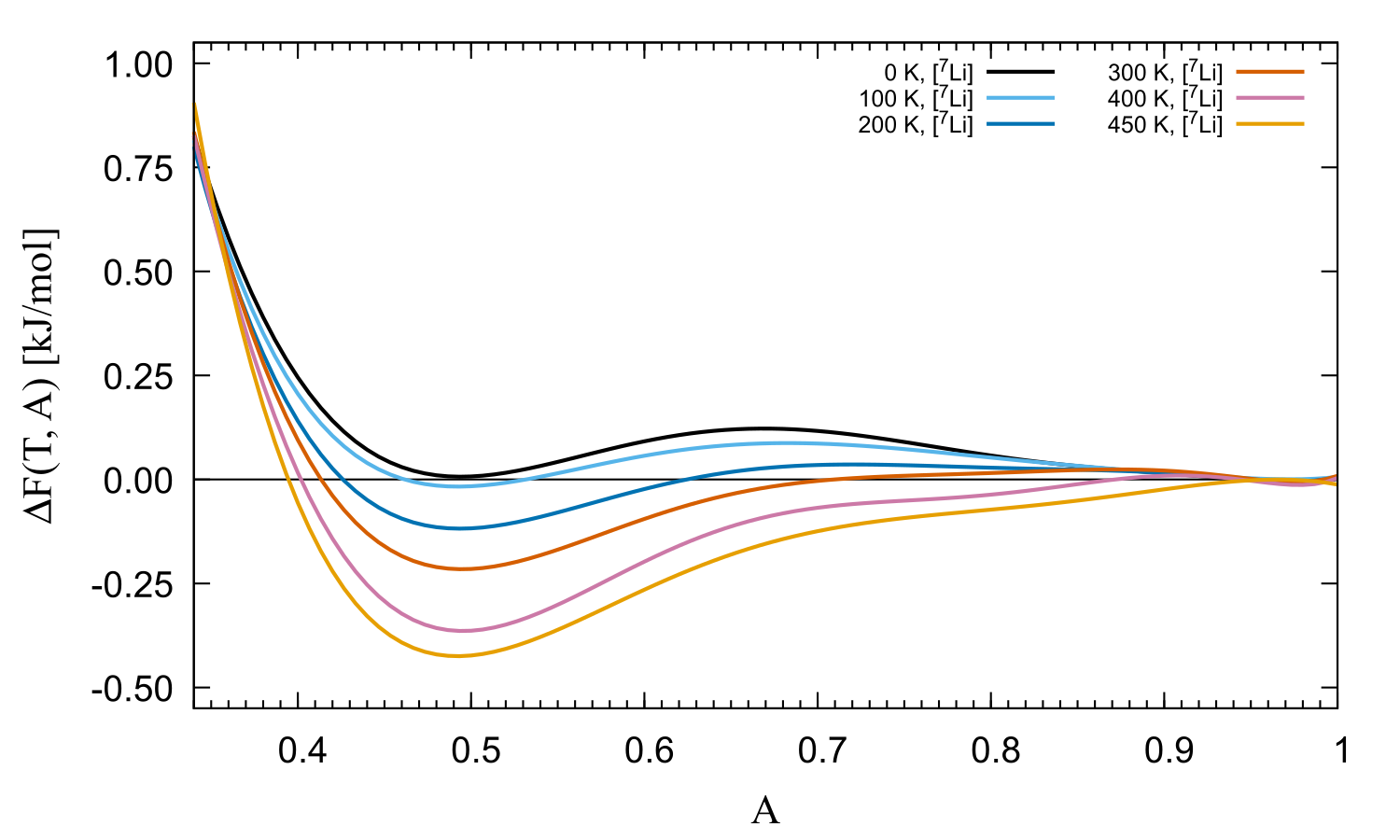}
  \caption{Differences in Helmholtz free energies $\Delta F(T,A) = F(T,A) - F(T,\mathrm{\textit{fcc}})$ at the PBE level of theory. The 0~K-curve contains $E_{\mathrm{coh}}$ and ZPVE terms. All results shown are at the optimized $A$-dependent structure obtained from the internal (cohesive) DFT energy only.}
  \label{fig:rel_free-energy-vs-a}
\end{figure}

Fig. \ref{fig:rel_free-energy-vs-a} shows results of $\Delta F(T,A) = F(T,A) - F(T,\mathrm{\textit{fcc}})$ for various temperatures up to 450~K close to the melting point (453.7~K\autocite{schaeffer2012a}). We see a rather interesting behaviour emerging for the cuboidal transformation. Increasing the temperature, the \textit{bcc}$\rightarrow$\textit{fcc} activation barrier increases substantially along the cuboidal path and moves towards the \textit{fcc} structure. Close to the melting point the \textit{fcc} structure becomes unstable with an almost barrierless transition into the \textit{bcc} structure. This is in line with the basic Landau picture where symmetry considerations favour the \textit{bcc} structure near the melting line.\autocite{alexander1978a} Interestingly, both the \textit{mcc} and \textit{acc} phases are not stable structures at any given temperature.

Analyzing the different contributions in Eq. \ref{eq:free-energy} we see the following trends emerging. The ZPVE contribution leads to a stabilization of the \textit{bcc} relative to \textit{fcc} phase (Fig. \ref{fig:e-vs-a_plot_rel-min_pbe_zpe}), which can be understood from the smaller coordination (kissing) number CN=8 compared to the \textit{fcc} structure with CN=12 leading to a reduced ZPVE for the \textit{bcc} phase. With the same reasoning, adding vibrational thermal contributions, $E_{\mathrm{therm}}$ in Eq. \ref{eq:free-energy}, further stabilizes the \textit{bcc} structure and shifts the minimum slightly in direction of \textit{acc}. Concerning entropy effects, we find for all temperatures $S^{\mathrm{\textit{bcc}}}_{\mathrm{vib}}>S^{\mathrm{\textit{mcc}}}_{\mathrm{vib}}>S^{\mathrm{\textit{acc}}}_{\mathrm{vib}}>S^{\mathrm{\textit{fcc}}}_{\mathrm{vib}}$ and therefore the term $-T \Delta S_{\mathrm{vib}}=-TS^{\mathrm{\textit{bcc}}}_{\mathrm{vib}}+TS^{\mathrm{\textit{fcc}}}_{\mathrm{vib}}<0$ in Eq. \ref{eq:free-energy} stabilizes the \textit{bcc} structure and, as the entropy term becomes dominant at higher temperatures, shifts the minimum slightly toward \textit{acc}.

We note that our data are in generally good agreement Hutcheon and Needs\autocite{hutcheon2019a} for the \textit{bcc} and \textit{fcc} structures, however, the cross-over point when the \textit{bcc} phase becomes more stable than the \textit{fcc} phase is critically dependent on the approximation applied. In our case, the \textit{bcc} phase becomes more stable than the \textit{fcc} phase at 49~K for \ce{^6Li} and at 64~K for \ce{^7Li}. While at higher temperatures, the \textit{fcc}$\rightarrow$\textit{bcc} transformation becomes energetically favoured through a barrierless transition, at low temperatures we have to consider a second path leading to the \textit{hcp} or Barlow-like multi-lattice arrangements,\autocite{ackland2017a,caspersen2005a} which is currently under investigation in our groups.

In conclusion, we have shown by analyzing the cuboidal martensitic transformation path,\autocite{burrows2021a} that phonons are responsible for the energetic quasi-degeneracy between the \textit{bcc} and \textit{fcc} lattice, in agreement with an earlier study by Faglioni et al.\autocite{faglioni2016a} Rather low barriers are involved in the martensitic phase transformation. Such tiny energy differences in the range of a few 0.1~kJ/mol are not so easily picked up in molecular dynamics or Monte-Carlo simulations and requires special care in the density functional and phonon dispersion calculations. Higher temperatures further stabilize the \textit{bcc} compared to the \textit{fcc} structure with an almost barrierless \textit{fcc}$\rightarrow$\textit{bcc} transition. Near the melting line the \textit{bcc} structure is often observed in \textit{acc}ordance with Landau theory,\autocite{anderson1985a,schaeffer2012a,alexander1978a} however, the \textit{bcc} and \textit{fcc} structures are quasi-degenerate, and near the melting point one has most likely a rapid conversion between \textit{bcc}, \textit{fcc}, \textit{hcp} and other cuboidal structures. Because of the kinetic hindrance at low temperature one can stabilize several phases and this may explain the controversy in the literature.\autocite{ackland2017a} For a more complete picture one requires in addition the volume/pressure dependence, exploring other martensitic transformations through molecular dynamics, and a higher level of theory such as the random phase approximation or quantum Monte Carlo, which currently is computationally too demanding.

  \printbibliography

@article{wang2015a,
  citation-number = {1},
  author = {Wang, Y. and Liu, B. and Li, Q. and Cartmell, S. and Ferrara, S. and Deng, Z.D. and Xiao, J.},
  year = {2015},
  volume = {286},
  pages = {330–345},
  journal = {J. Power Source}
}

@article{overhauser1984a,
  citation-number = {2},
  author = {Overhauser, A.W.},
  year = {1984},
  volume = {53},
  pages = {64–65},
  journal = {Phys. Rev. Lett.}
}

@article{smith1987a,
  citation-number = {3},
  author = {Smith, H.G.},
  year = {1987},
  volume = {58},
  pages = {1228–1231},
  journal = {Phys. Rev. Lett.}
}

@article{ashcroft1989a,
  citation-number = {4},
  author = {Ashcroft, N.W.},
  year = {1989},
  volume = {39},
  pages = {10552–10558},
  journal = {Phys. Rev. B}
}

@article{schwarz1990a,
  citation-number = {5},
  author = {Schwarz, W. and Blaschko, O.},
  year = {1990},
  volume = {65},
  pages = {3144–3147},
  journal = {Phys. Rev. Lett.}
}

@article{ackland2017a,
  citation-number = {6},
  author = {Ackland, G.J. and Dunuwille, M. and Martinez-Canales, M. and Loa, I. and Zhang, R. and Sinogeikin, S. and Cai, W. and Deemyad, S.},
  year = {2017},
  volume = {356},
  pages = {1254–1259},
  journal = {Science}
}

@article{smith1990a,
  citation-number = {7},
  author = {Smith, H.G. and Berliner, R. and Jorgensen, J.D. and Nielsen, M. and Trivisonno, J.},
  year = {1990},
  volume = {41},
  pages = {1231–1234},
  journal = {Phys Rev. B},
  number = {R}
}

@article{barlow1883a,
  citation-number = {8},
  author = {Barlow, W.},
  year = {1883},
  volume = {29},
  pages = {205–207},
  journal = {Nature},
  number = {739}
}

@article{berliner1986a,
  citation-number = {9},
  author = {Berliner, R. and Werner, S.A.},
  year = {1986},
  volume = {34},
  pages = {3586–3603},
  journal = {Phys. Rev. B}
}

@article{ashcroft2002a,
  citation-number = {10},
  author = {Ashcroft, N.W.},
  year = {2002},
  volume = {419},
  pages = {569–571},
  journal = {Nature},
  number = {6907}
}

@article{schaeffer2015a,
  citation-number = {11},
  author = {Schaeffer, A.M. and Temple, S.R. and Bishop, J.K. and Deemyad, S.},
  year = {2015},
  volume = {112},
  pages = {60–64},
  journal = {Proc. Nat. Acad. Sci.}
}

@article{struzhkin2002a,
  citation-number = {12},
  author = {Struzhkin, V.V. and Eremets, M.I. and Gan, W. and Mao, H.-K. and Hemley, R.J.},
  year = {2002},
  volume = {298},
  pages = {1213–1215},
  journal = {Science}
}

@article{shimizu2002a,
  citation-number = {13},
  author = {Shimizu, K. and Ishikawa, H. and Takao, D. and Yagi, T. and Amaya, K.},
  year = {2002},
  volume = {419},
  journal = {Nature},
  pages = {597–599}
}

@article{guillaume2011a,
  citation-number = {14},
  author = {Guillaume, C.L. and Gregoryanz, E. and Degtyareva, O. and McMahon, M.I. and Hanfland, M. and Evans, S. and Guthrie, M. and Sinogeikin, S.V. and Mao, H.-K.},
  year = {2011},
  volume = {7},
  pages = {211–214},
  journal = {Nature Phys.}
}

@article{matsuoka2009a,
  citation-number = {15},
  author = {Matsuoka, T. and Shimizu, K.},
  year = {2009},
  volume = {458},
  pages = {186–189},
  journal = {Nature}
}

@article{neaton1999a,
  citation-number = {16},
  author = {Neaton, J.B. and Ashcroft, N.W.},
  year = {1999},
  volume = {400},
  pages = {141–144},
  journal = {Nature}
}

@article{barakat1986a,
  citation-number = {17},
  author = {Barakat, B. and Bacis, R. and Carrot, F. and Churassy, S. and Crozet, P. and Martin, F. and Verges, J.},
  year = {1986},
  volume = {102},
  pages = {215–227},
  journal = {Chem. Phys.}
}

@article{burrows2021a,
  citation-number = {18},
  author = {Burrows, A. and Cooper, S. and Schwerdtfeger, P.},
  year = {2021},
  volume = {104},
  pages = {035306},
  journal = {Phys. Rev. E}
}

@article{kraft1993a,
  citation-number = {19},
  author = {Kraft, T. and Marcus, P.M. and Methfessel, M. and Scheffler, M.},
  year = {1993},
  volume = {48},
  pages = {5886–5890},
  journal = {Phys. Rev. B}
}

@article{rollmann2007a,
  citation-number = {20},
  author = {Rollmann, G. and Gruner, M.E. and Hucht, A. and Meyer, R. and Entel, P. and Tiago, M.L. and Chelikowsky, J. R.},
  year = {2007},
  volume = {99},
  pages = {083402–1–4},
  journal = {Phys. Rev. Lett}
}

@article{cayron2015a,
  citation-number = {21},
  author = {Cayron, C.},
  year = {2015},
  volume = {96},
  pages = {189–202},
  journal = {Acta Materialia}
}

@article{pichl1999a,
  citation-number = {22},
  author = {Pichl, W. and Krystian, M.},
  year = {1999},
  volume = {A273–275},
  pages = {208–212},
  journal = {Mat. Sci. Eng.}
}

@article{hecht2020a,
  citation-number = {23},
  author = {Hecht, U. and Gein, S. and Stryzhyboroda, O. and Eshed, E. and Osovski, S.},
  year = {2020},
  volume = {7},
  pages = {287–1–11},
  journal = {Front. Mat.}
}

@article{barrett1956a,
  citation-number = {24},
  author = {Barrett, C.S.},
  year = {1956},
  volume = {9},
  pages = {671–677},
  journal = {Acta Cryst.}
}

@article{vaks1989a,
  citation-number = {25},
  author = {Vaks, V.G. and Katsnelson, M.I. and Koreshkov, V.G. and Likhtenstein, A.I. and Parfenov, O.E. and Skok, V.F. and Sukhoparov, V.A. and Trefilov, A.V. and Chernyshov, A.A.},
  year = {1989},
  volume = {1},
  pages = {5319–5335},
  journal = {J. Phys.: Condens. Matter}
}

@article{ahlers2004a,
  citation-number = {26},
  author = {Ahlers, M.},
  year = {2004},
  volume = {9},
  pages = {169–183},
  journal = {Revista Matéria}
}

@article{schwarz1991a,
  citation-number = {27},
  author = {Schwarz, W. and Blaschko, O. and Gorgas, I.},
  year = {1991},
  volume = {44},
  pages = {6785–6790},
  journal = {Phys. Rev. B}
}

@article{conway1994a,
  citation-number = {28},
  author = {Conway, J. and Sloane, N.},
  year = {1994},
  volume = {48},
  pages = {373–382},
  journal = {J. Number Theory}
}

@article{bain1924a,
  citation-number = {29},
  author = {Bain, E.C.},
  year = {1924},
  volume = {16},
  pages = {692–698},
  journal = {Industrial \& Engineering Chemistry}
}

@article{doll1999a,
  citation-number = {30},
  author = {Doll, K. and Harrison, N.M. and Saunders, V.R.},
  year = {1999},
  volume = {11},
  pages = {5007–5019},
  journal = {J. Phys.: Cond. Matter}
}

@article{faglioni2016a,
  citation-number = {31},
  author = {Faglioni, F. and Merinov, B.V. and Goddard, W.A.},
  year = {2016},
  volume = {120},
  pages = {27104–27108},
  journal = {J. Phys. Chem. C}
}

@article{felice1977a,
  citation-number = {32},
  author = {Felice, R.A. and Trivisonno, J. and Schuele, D.E.},
  year = {1977},
  volume = {16},
  pages = {5173–5184},
  journal = {Phys. Rev. B}
}

@article{anderson1985a,
  citation-number = {33},
  author = {Anderson, M.S. and Swenson, C.A.},
  year = {1985},
  volume = {31},
  pages = {668–680},
  journal = {Phys. Rev. B}
}

@article{yao1996a,
  citation-number = {34},
  author = {Yao, G. and Xu, J.G. and Wang, X.W.},
  year = {1996},
  volume = {54},
  pages = {8393–8397},
  journal = {Phys. Rev. B}
}

@article{hopkins2011a,
  citation-number = {35},
  author = {Hopkins, A.B. and Stillinger, F.H. and Torquato, S.},
  year = {2011},
  volume = {83},
  pages = {011304–1–19},
  journal = {Phys. Rev. E}
}

@article{koumatos2016a,
  citation-number = {36},
  author = {Koumatos, K. and Muehlemann, A.},
  year = {2016},
  volume = {472},
  pages = {20150865},
  journal = {Proc. Royal Soc. A: Math. Phys. Eng. Sci.}
}

@article{li2017a,
  citation-number = {37},
  author = {Li, B. and Qian, G. and Oganov, A.R. and Boulfelfel, S.E. and Faller, R.},
  year = {2017},
  volume = {146},
  pages = {214502–1–7},
  journal = {J. Chem. Phys.}
}

@article{dunlap2017a,
  citation-number = {38},
  author = {Dunlap, R.A.},
  year = {2017},
  volume = {3},
  pages = {17–4},
  journal = {Eur. J. Phys. Ed.}
}

@article{xie2008a,
  citation-number = {39},
  author = {Xie, Y. and Ma, Y.M. and Cui, T. and Li, Y. and Qiu, J. and Zou, G.T.},
  year = {2008},
  volume = {10},
  pages = {063022},
  journal = {New J. Phys.}
}

@article{liu1999a,
  citation-number = {40},
  author = {Liu, A.Y. and Quong, A.A. and Freericks, J.K. and Nicol, E.J. and Jones, E.C.},
  year = {1999},
  volume = {59},
  pages = {4028–4035},
  journal = {Phys. Rev. B}
}

@article{hanfland2000a,
  citation-number = {41},
  author = {Hanfland, M. and Syassen, K. and Christensen, N.N.E. and Novikov, D.L.},
  year = {2000},
  volume = {408},
  pages = {174–178},
  journal = {Nature}
}

@article{karasiev2012a,
  citation-number = {42},
  author = {Karasiev, V.V. and Sjostrom, T. and Trickey, S.B.},
  year = {2012},
  volume = {86},
  pages = {056704–1–12},
  journal = {Phys. Rev. E}
}

@article{hutcheon2019a,
  citation-number = {43},
  author = {Hutcheon, M. and Needs, R.},
  year = {2019},
  volume = {99},
  pages = {014111–1–7},
  journal = {Phys. Rev. B}
}

@article{schaeffer2012a,
  citation-number = {44},
  author = {Schaeffer, A.M.J. and Talmadge, W.B. and Temple, S.R. and Deemyad, S.},
  year = {2012},
  volume = {109},
  pages = {185702–1–5},
  journal = {Phys. Rev. Lett.}
}

@article{alexander1978a,
  citation-number = {45},
  author = {Alexander, S. and McTague, J.},
  year = {1978},
  volume = {41},
  pages = {702–705},
  journal = {Phys. Rev. Lett.}
}

@article{caspersen2005a,
  citation-number = {46},
  author = {Caspersen, K.J. and Carter, E.A.},
  year = {2005},
  volume = {102},
  pages = {6738–6743},
  journal = {Proc. Nat. Acad. Sci.}
}






\end{document}